\documentclass[12pt]{article}
\title{On the renormalization of the Gibbons-Hawking boundary term}
\author{}
\newcommand{\lanln}[1]{$\langle$\texttt{arXiv:#1}$\rangle$}
\usepackage{verbatim}
\usepackage{color}
\usepackage{amsfonts}
\usepackage{amsmath}
\usepackage{amssymb}
\usepackage{latexsym}
\usepackage{psfrag}
\usepackage{lastpage}
\usepackage[pdftex]{graphicx} 
\usepackage[center]{caption2}

\date{}
\begin{document}

\begin{center} 
{\Large \bf On the renormalization of the \\ \vskip 2mm Gibbons-Hawking boundary term}
 \end{center}

\vskip 5mm
\begin{center} \large
{{Ted Jacobson\footnote{jacobson@umd.edu} and Alejandro Satz\footnote{alesatz@umd.edu}
}}
\end{center}

\vskip  0.5 cm
{\centerline{\it Maryland Center for Fundamental Physics}}
{\centerline{\it Department of Physics, University of Maryland}}
{\centerline{\it College Park, MD 20742-4111, USA}
}

\vskip 1cm

\begin{abstract}\small
{
The bulk (Einstein-Hilbert) and boundary (Gibbons-Hawking) terms in the gravitational action are generally renormalized differently when integrating out quantum fluctuations. The former is affected by nonminimal couplings, while the latter is affected by boundary conditions. We use the heat kernel method to analyze this behavior for a nonminimally coupled scalar field, the Maxwell field, and the graviton field. Allowing for Robin boundary conditions, we examine in which cases the renormalization preserves the ratio of boundary and bulk terms required for the effective action to possess a stationary point. The implications for field theory and black hole entropy computations are discussed.}
\end{abstract}

\section{Introduction}

Quantum field fluctuations, when integrated out on a curved background spacetime, 
generally lead to renormalization of gravitational couplings. If the background
has a boundary, then also the couplings in the boundary action are 
renormalized. The boundary conditions on the quantum fields do not affect the local bulk couplings but can affect the local boundary couplings; on the other hand, non-minimal coupling terms in the action affect the bulk but not the boundary couplings \cite{HKUM}. 
Renormalization can therefore change the balance between bulk and boundary 
terms. 

In this paper we investigate the impact of non-minimal couplings and boundary conditions 
for matter and metric fields on  
the renormalization of the bulk Einstein-Hilbert (EH) term ($\int_{\cal M} \sqrt{g} R$)
and the boundary Gibbons-Hawking (GH) term ($\int_{\partial\cal M} \sqrt{h} K$).
On a manifold with boundary, the GH term must appear
with the relative coefficient $2$ if the metric variation of the total action is to
have no boundary term when the boundary metric is held fixed. This
is required in order for the action to be stationary at solutions to the equations of motion.
The same is true in the presence of non-minimally coupled matter fields, when
the Ricci scalar appears multiplied by a function of fields in the action.
If this bulk/boundary balance were to be upset, it seems that the existence of a classical
limit and a valid perturbative expansion around a stationary point would be compromised.

In particular, when Dirichlet boundary conditions are used for a non-minimally coupled 
 field, the bulk and boundary term of the $a_2$ coefficient of the heat kernel expansion (which provides the renormalization of Newton's constant) do not have the proper balance \cite{Barv-Solod,reuterbecker}. This was interpreted in \cite{Barv-Solod} as indicating that the boundary renormalization should
be evaluated as a limit from the bulk, restoring the balance, while it was argued in \cite{reuterbecker} that the non-balanced result is actually the correct one.  Within the exact renormalization group approach that the latter authors use, this implies that the beta functions for the ``bulk Newton constant'' and the ``boundary Newton constant'' are different, which causes the scale-dependent effective action $\Gamma_k$ to be ``mismatched'' and not lead to proper equations of motion (except at most at one scale where the two couplings can be fixed as equal as an RG initial condition).

The Gibbons-Hawking term also plays an important role 
in black hole thermodynamics. When the Euclidean path integral for matter and gravitational fields 
is evaluated on a black hole solution that is a stationary point of the effective gravitational action, the (renormalized) entropy and other thermodynamical quantities are computed only from this boundary term. This is sometimes called the ``on-shell'' method for computing black hole entropy and its quantum corrections \cite{gibbonshawk, wilzec, reg1}. 
Thus it would seem that  the choice of boundary conditions
could affect the (leading order) value of the entropy when computed using the on shell method.
If so, it would lead to a puzzle, because
the entropy can also be computed from an ``off-shell'' procedure involving a 
conical defect \cite{suss2, solod2},
in which case the entropy depends only upon the bulk action. 
Thus it would seem that the two methods would generally not yield the same results
for all boundary conditions.

Questions about the equations of motion and the black hole entropy
are not directly physical when phrased in terms of regulated, would-be divergent
quantities. One might therefore take the viewpoint that only the renormalized effective action
need exhibit a proper matching of bulk and boundary couplings. However, renormalization
also contributes finite observable effects, such as under a change of renormalization scale as 
mentioned above or, more specifically, when integrating out the effects of a particular
massive field. Such questions provide the motivation for this paper, but they will not be fully resolved here.
We aim to develop some results that should contribute toward understanding them.
In particular, we shall study the renormalization of the EH and GH terms under
a broader class of boundary conditions than has been previously studied in this context,
including Robin conditions of a particular kind. For simplicity, we will restrict ourselves to the one-loop effective action framework assuming non-interacting quantum fields, but we expect our conclusions to be transferable to the exact renormalization group approach. 

In section 2, we summarize the computation of the effective action from the heat kernel expansion and its dependence on non-minimal coupling and boundary conditions. In section 3, we discuss the case of the non-minimally coupled scalar field, and we find that a particular Robin boundary condition is the only one ensuring the balanced renormalization of the GH boundary term. (The necessity of Robin boundary conditions for preserving
bulk/boundary balance in the presence
of non-minimal coupling has previously been argued by 
Solodukhin \cite{Soloddraft},
but to our knowledge it is not discussed in the literature.) In section 4 we discuss the Maxwell field, and find
that in this case the boundary conditions compatible with gauge invariance, which are of mixed Dirichlet-Robin type, automatically ensure the balance. In section 5 we discuss the case of the graviton field. 
A naive analysis shows that the balance is not preserved when using gauge-invariant boundary conditions (on certain backgrounds); however, we argue that the question for gravitons is more subtle and requires further study.
Section 6 includes a discussion of the main results and the questions they leave open, and a discussion and resolution of the puzzle about black hole entropy mentioned above.

\section{Heat kernel and boundary conditions}

We consider a set of free quantum fields $\varphi^A$, labelled by a tensor, spinor  or internal index $A$, on a 4-dimensional manifold $\mathcal{M}$ with background Euclidean metric $g_{\mu\nu}$ and boundary $\partial \mathcal{M}$. We assume that the action can be written in the form:
\begin{align}\label{action}
S[g,\varphi^A]&=-\frac{1}{16\pi G_b}\int_\mathcal{M} \mathrm{d}^4 x \sqrt{g}\, (R\,-2\Lambda_b\,+\,\cdots)\,+\,\frac{1}{2}\int_\mathcal{M} \mathrm{d}^4 x \,\sqrt{g}\, \varphi^A\,D_{AB}\,\varphi^B\nonumber\\
&+S_\partial[g, \varphi^A]\,,
\end{align}
where we have included a bare gravitational action (the dots stand for unwritten higher-order in curvature terms). Here $D_{AB}$ is a second-order differential operator of the Laplace type, by which we mean that it has the structure:
\begin{equation}\label{laplace}
D_{AB} = -(\eta_{AB} g^{\mu \nu}\nabla_\mu\nabla_\nu + E_{AB})\,,
\end{equation}
where $\nabla$ is the  covariant derivative, $\eta_{AB}$ a suitable metric\footnote{For a collection of scalars, $\eta_{AB}$ is the identity matrix. For a vector field it is the spacetime metric, and for a tensor field the DeWitt metric.} on the configuration space $\varphi^A$, and $E_{AB}$ involves no derivatives and can include mass parameters as well as background structures such as the curvature. (In this, as in most of this section, we follow the notation of \cite{HKUM}.)  $S_\partial$ in (\ref{action}) is a suitable boundary action that leads to well-defined equations of motion when $S$ is varied with respect to both $g_{\mu\nu}$ and $\varphi^A$; neglecting higher order curvature terms, its purely gravitational part consists of the well-known Gibbons-Hawking term,
\begin{equation}
-\frac{1}{8\pi G_b}\int_{\partial \mathcal{M}}\mathrm{d}^3x\sqrt{h}\,K\,,
\end{equation}
where $h$ is the induced metric on $\partial \mathcal{M}$ and $K$ the extrinsic curvature.\footnote{The extrinsic curvature is defined as $K = h^{ij}K_{ij}$, where $K_{ij}=-\frac{1}{2}\mathcal{L}_n\,h_{ij}$ defines the second fundamental form by the Lie derivative of the intrinsic boundary metric along the outward normal direction. The reason for the minus sign is that we denote by $n^{\mu}$ the \textit{inward}-directed unit normal. This follows the conventions of \cite{HKUM}. Note that under these conventions $K>0$ for a sphere.} This might be supplemented by terms similarly required for ``balancing'' the matter action and leading to well-defined equations of motion upon variation subject to the boundary conditions.

 For example, for a single massless nonminimally coupled scalar field $\varphi$ we may have $\eta_{AB}=1$, $E_{AB}=-\xi\,R$, and
\begin{displaymath}
S_\partial[g,\varphi]=\int_{\partial \mathcal{M}}\mathrm{d}^3x\sqrt{h}\,K\left[-\frac{1}{8\pi G_b}+\xi\,\varphi^2\right]\,,
\end{displaymath}
Upon variation with respect to  $g_{\mu\nu}$ and $\varphi$, subject to Dirichlet boundary conditions on the metric ($\delta g_{\mu\nu}\big|_{\partial\mathcal{M}}=0$) and the scalar field ($\varphi\big|_{\partial\mathcal{M}}=0$), the boundary term in the variation vanishes and the bulk one yields the equations of motion: the Einstein equations with source $T_{\mu\nu}[\varphi]$, and the Klein-Gordon equation with nonminimal coupling $\xi$. The  addition of a suitable extra term to the boundary action can allow the Dirichlet boundary condition to be replaced by a Robin boundary condition, as described in detail in Section 3.

The quantum theory of $\varphi^A$ on a classical background $g_{\mu\nu}$, is defined by the path integral 
\begin{equation}
Z[g]=\int\mathcal{D}\varphi^A\,\mathrm{e}^{-S[g,\varphi^A]}\,,
\end{equation}
where the integral is done over fields satisfying suitable boundary conditions at $\partial \mathcal{M}$. Expanding the action as in (\ref{action}) into bulk and boundary parts, it is seen that if the boundary action $S_\partial$ vanishes for fields satisfying the boundary conditions, the boundary action makes no contribution and the path integral evaluates to the determinant of the operator $D$ (up to an irrelevant constant).\footnote{We discuss a different viewpoint at the end of Section 3.} Defining the effective action $\Gamma[g] = -\ln Z$, we have then at the formal level:
\begin{equation}\label{gamma}
\Gamma[g]=S_b[g]+\frac{1}{2}\,\mathrm{Tr} \ln D\,\equiv S_b[g]+W[g].
\end{equation}

The $W$-term is divergent and needs regularization.\footnote{We should also introduce for dimensional consistency a mass scale $\mu$ in the path integral measure, leading to $\mathrm{Tr}\ln\left(D/\mu^2\right)$ in (\ref{gamma}). The $\mu$-dependence in $\Gamma$ can be cancelled as well by adjusting the bare action and plays no role in the rest of our analysis.} We will define it through the heat kernel expansion with a short-distance cutoff $\epsilon$, and cancel the terms divergent as $\epsilon\to 0$ by suitable counter terms in the bare action $S_b$, leading to renormalized gravitational couplings in $\Gamma$. In this framework, we can write
\begin{equation}\label{WandK}
W[g]=-\frac{1}{2}\int_{\epsilon^2}^\infty\mathrm{d}t\,\frac{K(t,D)}{t}\,,
\end{equation}
where the trace of the heat kernel, $K(t,D)$, is given by
\begin{equation}
K(t,D)=\mathrm{Tr} \left[\mathrm{e}^{-t\,D}\right]\,.
\end{equation}
$K(t,D)$ has a small $t$ expansion in terms of the form
\begin{equation}\label{Kexpansion}
K(t,D)\sim \sum_{j\geq 0}t^{(j-4)/2}a_j(D)\,,
\end{equation} 
where the heat kernel coefficients $a_k$ are integrals over $\mathcal{M}$ and $\partial\mathcal{M}$ involving geometrical tensors, the matrix $E$, and quantities appearing in  in the expression of the boundary conditions.

From (\ref{WandK}) and (\ref{Kexpansion}) it follows that the leading divergences in $W$ can be expressed as
\begin{equation}\label{Wexpansion}
W[g]\sim -\frac{1}{2}\left\{ \frac{1}{2\epsilon^4}\,a_0(D)+\frac{2}{3\epsilon^3}\,a_1(D)+\frac{1}{\epsilon^2}\,a_2(D)+\cdots\right\}\,.
\end{equation}
Let us assume that the boundary conditions imposed are either Dirichlet,
\begin{equation}
\varphi^A\big{|}_{\partial\mathcal{M}} = 0\,,
\end{equation}
or Robin,
\begin{equation}\label{robindef}
\left(n^\mu\nabla_\mu\,\varphi^A+S_{AB}\,\varphi^B\right)\big{|}_{\partial\mathcal{M}} = 0\,,
\end{equation}
where $n^{\mu}$ is the inwards-pointing normal unit vector on $\partial\mathcal{M}$; the particular case of Neumann boundary conditions is covered when $S=0$. The heat kernel coefficients corresponding to these boundary conditions have been computed in \cite{HKUM, mckean,dowker} and are:
\begin{equation}
a_0(D)=\frac{1}{(4\pi)^2}\int_{\mathcal{M}}\mathrm{d}^4x\,\sqrt{g}\,d\,,
\end{equation}
\begin{equation}\label{a1}
a_1(D)=\pm\frac{1}{4}\frac{1}{(4\pi)^{3/2}}\int_{\partial\mathcal{M}}\mathrm{d}^3x\,\sqrt{h}\,d\,,
\end{equation}
\begin{equation}\label{a2}
a_2(D)=\frac{1}{6}\frac{1}{(4\pi)^2}\left\{\int_{\mathcal{M}}\mathrm{d}^4x\,\sqrt{g}\,\left(R\,d+6 E^A_{\,\,A} \right)  +2\int_{\partial\mathcal{M}}\mathrm{d}^3x\,\sqrt{h}\left(K\,d+6 S^A_{\,\,A}\right)
\right\}\,.
\end{equation}

Here $d=\eta^{AB}\eta_{AB}$ stands for the dimension of the vector space where $\varphi^A$ lives, the $\pm$ in (\ref{a1}) stands for Robin and Dirichlet boundary conditions respectively, and $S_{AB}$  should be set to zero in (\ref{a2}) for the Dirichlet case. The case of mixed boundary conditions, when some components of $\varphi^A$ satisfy Dirichlet conditions and some satisfy Robin conditions, is covered by the same formulas, with the trace of $S_{AB}$  in (\ref{a2}) taken only over the subspace where it is defined.

We assume that the $\Lambda_b$ is set so that the combination $\Lambda_b/G_b$ in the bare gravitational action contains suitable divergences leading to a finite renormalized cosmological term in $\Gamma$. We further assume that $S_\partial$ contains an $\sim\epsilon^{-3}$ volume term that cancels $a_1$. Our attention in the rest of the paper will be focused on the $a_2$ term, and the renormalization of Newton's constant it implies for both the bulk and the boundary terms in the action.

It is immediate from (\ref{a2}) that if $E_{AB}$ does not involve $R$ (minimal coupling) and also the boundary condition coefficients $S_{AB}$ do not involve the extrinsic curvature $K$, then the bulk $R$ term and the boundary $K$ terms of $a_2$ have the same relation as the bulk $R$ term and boundary Gibbons-Hawking term in the bare gravitational action. This means that the renormalized effective action $\Gamma$ will also have the general form
\begin{equation}
\Gamma[g]=-\frac{1}{16\pi G_0}\int_\mathcal{M} \mathrm{d}^4 x \sqrt{g}\, (R\,-2\Lambda_0\,+\,\cdots)\,-\frac{1}{8\pi G_0}\int_{\partial \mathcal{M}}\mathrm{d}^3x\sqrt{h}\,K\,,
\end{equation}
in terms of renormalized couplings $\Lambda_0, G_0$; in particular, the renormalization of $G$ is written:
\begin{equation}
G_0 = G_b + \frac{d}{12\pi\epsilon^2}\,.
\end{equation}
This equation applies equally for the bulk term and for the boundary term of the action; this fact guarantees that $\delta\Gamma/\delta g = 0$ leads to the Einstein field equations, whereas if this bulk-boundary balance were broken, $\delta\Gamma$ would contain a boundary term with normal derivatives of $\delta g_{\mu\nu}$.

It is noteworthy that the heat kernel coefficients for a minimally coupled scalar field with Dirichlet or Neumann boundary conditions have precisely the ratio required in order to preserve the bulk-boundary balance for the renormalized gravitational couplings. To our knowledge, this fact has not been remarked upon before. It may be purely coincidental, but perhaps it can receive an explanation by considering how the heat kernel responds to variations of the background metric. If so, that might provide insight into what happens for nonminimally coupled fields and alternative boundary conditions, which will occupy our attention for most of this paper. 

Returning to (\ref{a2}), it is also clear that if the field is nonminimally coupled to the curvature (so that $E_{AB}$ includes $R$), then the Dirichlet boundary condition fails to produce the proper balance within the bulk and boundary terms of $a_2$. The same happens for minimally coupled fields if Robin boundary conditions are imposed with $S_{AB}$ involving $K$, or for nonminimally coupled fields if Robin boundary conditions are imposed that do not involve $K$ in $S_{AB}$ in a very particular way. This failure to achieve bulk-boundary balance would entail that a key property of the action (yielding the equations of motion when varied under Dirichlet conditions for the metric) is not preserved from the bare to the renormalized gravitational action. 

The observation concerning Dirichlet conditions and nonminimal coupling was made in \cite{reuterbecker}, in the context of the exact renormalization group, which we proceed to explain now in a brief detour from our main focus. When an effective action at scale $k$ is introduced, interpolating between the full effective action $\Gamma$ at $k\to 0$ and the bare action at $k\to\infty$, it satisfies the exact renormalization group equation
\begin{equation}\label{erge}
k\partial_k \Gamma_k = \frac{1}{2}\mathrm{Tr}\left[\frac{k\partial_k \mathcal{R}_k}{D_k+\mathcal{R}_k}\right]\,,
\end{equation}
where $D_k$ is the Hessian of $\Gamma_k$ with respect to the fields it depends on (thus being analogous to $D$ in our formalism), and $\mathcal{R}_k$ is an IR cutoff function which also depends on $D_k$. The trace of a general function $F(D_k)$ can be computed with the heat kernel method \cite{codello} as
\begin{equation}
\mathrm{Tr}[F(D_k)]=\sum_{j\geq 0}a_j(D_k)\,Q_{\frac{4-j}{2}}(F)\,,\quad\quad\, Q_n(F)=\frac{1}{\Gamma(n)}\int\mathrm{d}z\, z^{n-1}F(z)\,.
\end{equation}
In this case where $F(D_k)=(D_k+\mathcal{R}_k(D_k))^{-1}k\partial_k\mathcal{R}_k(D_k)$, the coefficients $Q_n$ will depend only on $k$ and the functional form of the cutoff $\mathcal{R}_k$; the details of the form of the operator $D_k$, and the boundary conditions, influence only the heat kernel coefficients $a_k$, which are computed in the same way as for the one-loop effective action. 

Hence, when a $\Gamma_k$ suitable for quantum gravity is expanded in geometric terms, the right-hand side of the flow equation (\ref{erge}) will give different beta functions to the coefficients in the bulk $\int R$ and the boundary $\int K$ terms, as long as the relation between a nonminimal term in $D_k$ and the boundary conditions imposed is not of the particular form discussed above. In particular, for Dirichlet boundary conditions any nonminimal coupling in the action entails a different renormalization group flow for the ``bulk Newton constant'' and the ``boundary Newton constant''. Then the variation of $\Gamma_k$ with respect to the metric does not lead to well-defined equations of motion, except at most at one scale $k_0$ where the balance can be postulated as part of the initial condition for the RG flow.

In the following sections we will address in turn the scalar field, the Maxwell field and the graviton field, and discuss for each of them the prospects of balancing the bulk and boundary renormalizations by employing Robin boundary conditions that are ``matched'' to the nonminimal coupling in a particular way. 

\section{Scalar field}

In this section we consider a single massless non-minimally coupled scalar field. We take the action to be
\begin{equation}\label{phiaction}
S[g,\varphi]=\frac{1}{2}\int_{\mathcal{M}}\mathrm{d}^4x\,\sqrt{g}\,\varphi\left(-\nabla^2+\xi R\right)\varphi+\int_{\partial\mathcal{M}}\mathrm{d}^3x\,\sqrt{h}\,\left(\xi K\varphi^2+\alpha\,\varphi\,\nabla_n \varphi\right)\,.
\end{equation}
We have included an extra boundary term proportional to $\varphi\nabla_n\varphi$ for compatibility with Robin boundary conditions; $\alpha$ is an arbitrary numerical parameter and $\nabla_n = n^\mu\nabla_\mu$ is the (inwards) normal derivative. The boundary term $\xi K \varphi^2$ is required to ensure that variation with respect to the metric leaves no uncancelled boundary term involving $\nabla_n g$. 

Issues related to the choice of boundary conditions for the non-minimally coupled scalar field have
been considered previously in several contexts by Solodukhin. 
In particular, he considered a Robin boundary condition, 
both in the context of the ``brick-wall'' technique for computing black hole entropy \cite{Solod97}, 
and in a more general context for manifolds with boundaries \cite{Soloddraft}, where also
the necessity of Robin boundary conditions for preserving bulk/boundary balance in the presence
of non-minimal coupling was argued. 
The full range of possible boundary conditions and 
associated boundary terms was also discussed in \cite{solod98}.

Variation of the action with respect to $\varphi$ leads to
\begin{align}\label{deltaS}
 \delta S &=\int_{\mathcal{M}} \mathrm{d}^4x\, \sqrt{g} \, \delta\varphi\,(-\nabla^2+\xi R)\varphi \nonumber\\
 &+\int_{\partial \mathcal{M}}\mathrm{d}^3x\, \sqrt{h} \left\{\left[ \left(\alpha+\frac{1}{2}\right)\nabla_n\varphi\,+2\,\xi K\,\varphi\right]\delta\varphi+\left(\alpha-\frac{1}{2}\right)\varphi\,\nabla_n(\delta\varphi)\right\}\,.
\end{align}
To infer the equation of motion $(-\nabla^2+\xi R)\varphi=0$ from the bulk term, a boundary condition that makes the boundary term vanish is needed. If $\alpha=0$, only the vanishing Dirichlet boundary condition, 
\begin{equation}
\varphi\big|_{\partial\mathcal{M}}=0\,,
\end{equation}
ensures the vanishing of the boundary term in (\ref{deltaS}). 
For nonzero $\alpha$, the Robin condition
\begin{equation}\label{robin}
\left(\nabla_n\varphi+\frac{\xi K}{\alpha}\varphi\right)\Big|_{\partial\mathcal{M}}=0\,,
\end{equation}
is also permissible.
The non-vanishing Dirichlet condition can be implemented with the choice $\alpha=1/2$, and 
the Neumann boundary condition can be
implemented with the limit $\alpha\to\infty$.

According to  (\ref{Wexpansion}) and (\ref{a2}), the terms of $W$ that are linear in (bulk or extrinsic) curvature are
\begin{equation}\label{WRK}
W\big|_{R,K}=-\frac{1}{12\epsilon^2}\frac{1}{(4\pi)^2}\left\{ \int_{\mathcal{M}}\mathrm{d}^4x\,\sqrt{g}\,R\left(1-6 \xi \right)  +2\int_{\partial\mathcal{M}}\mathrm{d}^3x\,\sqrt{h}K\left(1+6 S\right) \right\}\,.
\end{equation}
Here $S$ should be set to zero for Dirichlet boundary conditions and to $\xi/\alpha$ for Robin boundary conditions of the form (\ref{robin}). It is then clear that the balance between the bulk and the boundary terms of the gravitational effective action can be preserved only if the boundary condition is Robin with the value $S=-\xi$, corresponding to the special value $\alpha = -1$

The renormalization of Newton's constant in both bulk and boundary terms reads for this value of $\alpha$:
\begin{equation}\label{Grenscalar}
\frac{1}{G_0}=\frac{1}{G_b}+\frac{1}{12\pi \epsilon^2}\left(1-6\xi \right)\,.
\end{equation}
Had we imposed different boundary conditions, we would have still this expression for the renormalization of $G_{\mathrm{bulk}}$ (i.e. the coefficient of the $\int R$ term) and a different one for the renormalization of $G^{(\partial)}$ (i.e. the coefficient of the boundary $K$ term):
\begin{equation}\label{Gbrenscalar}
\frac{1}{G^{(\partial)}_0}=\frac{1}{G^{(\partial)}_b}+\frac{1}{12\pi \epsilon^2}\left(1+\frac{6\xi}{\alpha} \right)\,.
\end{equation}
where $\alpha\to\infty$ covers both the Dirichlet and the Neumann case.

We reach therefore the conclusion that when a non-minimally coupled scalar is integrated out, the balanced renormalization of the bulk and boundary terms in the gravitational effective action (necessary in order to reproduce the Gibbons-Hawking term in $\Gamma[g]$) is not possible if the field satisfies Dirichlet boundary conditions, and is in fact only possible if it satisfies Robin boundary conditions of the specific form
\begin{equation}\label{robinpref}
\left(\nabla_n\varphi-\xi\,K\,\varphi\right)\big|_{\partial\mathcal{M}}=0\,.
\end{equation}
This is a surprising conclusion, because a wider range of boundary conditions (Dirichlet, Neumann, and Robin with arbitrary coefficients) are admissible for the scalar field on its own, and one could have naively expected that the effective dynamics of the gravitational field is well-defined for any of them.
Is there any way to avoid the conclusion that only one preferred 
Robin boundary condition is allowed for the non-minimally coupled scalar?

We will return to this issue in the Discussion section. For the moment we will limit ourselves to commenting on the different point of view taken in  \cite{Barv-Solod}, which also examines non-minimally coupled fields on manifolds with boundary. There the vanishing Dirichlet boundary condition is assumed, and hence the standard expression for the effective action $W= \frac{1}{2}\mathrm{Tr}\ln D$ exhibits bulk-boundary mismatch. However, the authors argue that the effective action contains in addition to this term a further one consisting of an integral over the boundary of $\xi K \langle \varphi^2\rangle$, evaluated as a limit approaching the boundary from the bulk. When this contribution is taken into account the bulk-boundary balance is restored. 

Explained in more detail, the argument in \cite{Barv-Solod} starts from the partition function for the action (\ref{phiaction}) with $\alpha=0$, and computes:
\begin{align}\label{barv}
W&=-\ln\int\mathcal{D}\varphi\,\mathrm{e}^{-\xi\int_{\partial\mathcal{M}}\mathrm{d}^3x\,\sqrt{h}\,K\,\varphi^2}\,\,\mathrm{e}^{-S_{\mathrm{bulk}}[g,\varphi]}\nonumber\\
&=\bar{W}+\xi\langle\int_{\partial\mathcal{M}} \mathrm{d}^3x\,\sqrt{h}\,K\,\varphi^2\rangle\,+O(\xi^2K^2)\,,
\end{align}
in a formal perturbative expansion; here $\bar{W}$ is the standard expression $ \frac{1}{2}\mathrm{Tr}\ln D$, where $D=S_{\mathrm{bulk}}^{(2)}=-\nabla^2+\xi R$. It is argued that despite the  Dirichlet boundary conditions, the expectation value $\langle\varphi^2\rangle$ does not vanish, and can be substituted by a standard heat kernel expansion evaluation. To the order we are interested in, it is computed from $a_0$ coefficient evaluated  as an integral over the boundary. Thus it is found that the one-loop contribution to the effective action is
\begin{equation}
W\big|_{R,K}=\bar{W}+\frac{1}{(4\pi\epsilon)^2}\xi\int_{\partial\mathcal{M}} \mathrm{d}^3x\,\sqrt{h}\,K\,+O(\xi^2K^2)\,.
\end{equation}
Comparison with the expression for $\bar{W}$ given by setting $S=0$ in (\ref{WRK}) shows that the bulk-boundary balance is exactly restored  by this procedure. 

However appealing this conclusion, we do not accept the reasoning that leads to it. The functional integral in the first line of (\ref{barv}) is supposed to be over all field configurations $\varphi(x)$ satisfying the chosen boundary conditions at $\partial\mathcal{M}$. If Dirchlet boundary conditions are chosen, it follows immediately that the boundary action vanishes and one must integrate only the exponential of the bulk action, a procedure leading back to $\bar{W}$. The calculation that obtains a separate boundary contribution, from this standpoint, appears to contain an illicit evaluation by a limiting procedure that first computes the expectation value $\langle\varphi^2\rangle$ in the bulk and then evaluates it in the limit that the boundary is approached.

\section{Maxwell field}

In this section we consider a gauge field $A_\mu$, restricting ourselves for simplicity to the Abelian case. The physics must be invariant under gauge transformations of the form
\begin{equation}\label{gaugetransf}
A_\mu\longrightarrow A'_\mu=A_\mu + \nabla_\mu\, \xi\,.
\end{equation}
The bulk action can be written as
\begin{equation}\label{Amuaction}
S_{\mathrm{bulk}}[g,A_\mu] = \frac{1}{2}\int_{\mathcal{M}}\mathrm{d}^4 x\,\sqrt{g}\,\left[-A_\mu \nabla^\nu\nabla_\nu A^\mu +A^\mu\nabla_\mu\nabla_\nu A^\nu+A_\mu A_\nu R^{\mu\nu}\right]\,.
\end{equation}

The quantum theory is defined by a path integral including Fadeev-Popov gauge fixing and ghost terms, which we also represent by $\xi$. We use the Lorenz gauge,
\begin{equation}
\nabla_\mu A^\mu = 0\,,
\end{equation}
which disposes of the second term in the square brackets in (\ref{Amuaction}) by adding its negative as the gauge fixing term. Then the full effective action including gauge and ghost contributions takes the form \cite{HKUM},
\begin{equation}\label{STR}
\Gamma[g]=S_b[g]+\frac{1}{2}\mathrm{Tr}\ln D^{(A)} - \mathrm{Tr}\ln D^{(\xi)}\,,
\end{equation}
with the gauge field and the ghost operators being
\begin{align}
D^{(A)}&=-g_{\mu\nu}\nabla^2 +R_{\mu\nu}\,\quad\quad d=4\,,\label{Maxop}\\
D^{(\xi)}&=-\nabla^2\,\,\,\quad\quad\quad\quad\quad\quad d=1\,.
\end{align}
We still need to fix the boundary conditions at $\partial{\mathcal{M}}$ which, as in the scalar case, can be done including a suitable boundary action so that the boundary term in $\delta S$ vanishes upon applying the boundary conditions. However, there is an extra requirement that must be met in the gauge field case. If one works  within the standard framework with Fadeev-Popov ghosts, the required gauge invariance of the physical results implies that boundary conditions must be invariant under the gauge transformations (\ref{gaugetransf}). In other words, if  $A_\mu$ satisfies the boundary conditions then $A'_\mu$  related by (\ref{gaugetransf}) must also \cite{HKUM,FPtrick}.

A Dirichlet boundary condition for both gauge and ghost fields, for example, does not satisfy this requirement; if we impose $A_\mu=0=\xi$ on $\partial{\mathcal{M}}$, then we will have that on  $\partial\mathcal{M}$
\begin{equation}
A'_n=\nabla_n\xi\neq 0\,.
\end{equation}
Working within the Lorenz gauge, there are two alternative sets of boundary conditions that, unlike the Dirichlet conditions, satisfy this requirement. They are called the ``absolute'' and the ``relative'' boundary conditions \cite{HKUM,Espositobook} and can be written respectively as:
\begin{align}
A_n\big|_{\partial\mathcal{M}}&=0\,,\quad\quad\quad (\nabla_n A_i - K_{ij}A^j)\big|_{\partial\mathcal{M}}=0\,,\nonumber\\
\nabla_n\xi\big|_{\partial\mathcal{M}}&=0\,,\label{absolute}
\end{align}
and
\begin{align}
A_i\big|_{\partial\mathcal{M}}&=0\,,\quad\quad\quad (\nabla_n A_n - K A_n)\big|_{\partial\mathcal{M}}=0\,,\nonumber\\
\xi\big|_{\partial\mathcal{M}}&=0\,,\label{relative}
\end{align}
where $i,j$ indices label tangential components and $n$ the normal component, and $K_{ij}$ is the second fundamental form on $\partial\mathcal{M}$ (whose trace is $K$),  and the indexed expressions denote the corresponding components of covariant tensors. It can be easily checked that both sets of boundary conditions are invariant under transformations (\ref{gaugetransf})\footnote{The invariance of the absolute boundary conditions is a matter of straightforward computation; that of the relative boundary conditions requires using that the modes summed over in the path integral are eigenfunctions of the ghost kinetic operator  $\nabla^2$, which implies that $\nabla^2\xi\big|_{\partial{\mathcal{M}}}\propto \xi\big|_{\partial{\mathcal{M}}}=0$ for them.}.
Both sets of boundary conditions are comptible with a boundary action added to (\ref{Amuaction}), and the boundary action itself vanishes for configurations satisfying the boundary conditions \cite{HKUM}.

The heat kernel coefficient $a_2$ is given by the same expression (\ref{a2}) with the only difference being that the trace of $S_{AB}$ is only taken over the subspace satisfying Robin conditions where $S$ is defined. For absolute boundary conditions we have that 
\begin{equation}
S^{\mathrm{\,\,\,abs}}_{AB}\longrightarrow -K_{ij}\,, 
\end{equation}
and for relative boundary conditions we have
\begin{equation}
S^{\mathrm{\,\,\,rel}}_{AB} \longrightarrow-K. 
\end{equation}
In both cases its trace is $-K$. The matrix $E$ equals $-R_{\mu\nu}$, its trace is $-R$, and it is clear that for both absolute and for relative boundary conditions the coefficient for the operator $D^{(A)}$ reads:
\begin{equation}
a_2(D^{(A)})=\frac{1}{6}\frac{1}{(4\pi)^2}\left\{\int_{\mathcal{M}}\mathrm{d}^4x\,\sqrt{g}\,R\,(+4-6)  +2\int_{\partial\mathcal{M}}\mathrm{d}^3x\,\sqrt{h}\,K\left(+4-6\right)
\right\}.
\end{equation}
where we have separated for clarity the contribution from the Laplacian operator ($+4R$ in bulk, $+4K$ in boundary) from that of the nonminimal coupling ($-6R$ in bulk) and that of the Robin boundary condition applied to certain components ($-6K$ in boundary). It is seen that the bulk and boundary terms have the appropriate balance. The coefficient for the ghost operator, which is simply the Laplacian subject to either Dirichlet or Neumann boundary conditions, is in both cases
\begin{equation}
a_2(D^{(\xi)})=\frac{1}{6}\frac{1}{(4\pi)^2}\left\{\int_{\mathcal{M}}\mathrm{d}^4x\,\sqrt{g}\,R +2\int_{\partial\mathcal{M}}\mathrm{d}^3x\,\sqrt{h}\,K
\right\}\,.
\end{equation}
Hence the total renormalization of Newton's constant due to integrating out the electromagnetic field applies equally to the bulk and boundary terms of the action, whether absolute or relative boundary conditions are chosen. Taking into account (\ref{STR}) for the relative weight of the ghost contribution, it reads:
\begin{equation}
\frac{1}{G_0} = \frac{1}{G_b}+\frac{1}{12\pi\epsilon^2}\left(4-6-2\right) = \frac{1}{G_b}-\frac{1}{3\pi\epsilon^2}\,.
\end{equation}

\section{Graviton field}

In this section we will discuss the renormalization of the Einstein-Hilbert action and the boundary Gibbons-Hawking term due to integrating out a tensor field $h_{\mu\nu}$, interpreted as a quantized perturbation of the background metric $g_{\mu\nu}$. One might think that the calculation is a simple generalization of the one for Maxwell fields; however, there are several subtleties and complications involved. 

The first one is that the linearized theory of a quantized tensor field $h_{\mu\nu}$  is gauge invariant only if 
the background is on-shell, that is, satisfies the bare Einstein equations \cite{deser}. However, our purposes require varying the effective action $\Gamma[g]$ with respect to the background metric $g_{\mu\nu}$ to see if the boundary term of the variation is cancelled (this it what the ``balanced renormalization'' amounts to). How is this variation to be carried out if $g_{\mu\nu}$ has been specified as an on-shell solution of the bare equations of motion? The proper answer to this question would presumably involve replacing the effective action for the background metric that we are using (suitable for analyzing the backreaction of quantum fields on curved spacetime) by the Legendre effective action, which depends on the expectation value of the metric field and is defined via the background field formalism \cite{abbot, clapper}. However, an exploration of the issue in the context of the background field method is beyond the scope of this paper. Here we restrict our attention to the bulk-boundary balance of the $a_2$ heat kernel coefficient of different kinds of Laplace operators (with different kinds of boundary conditions) on an arbitrary background. 
Although results of such calculations for gravitons are not directly applicable to our bulk/boundary balance question 
without further analysis, they could be useful in a more thorough treatment, as well as in other  applications. Hence we shall present them briefly here.

 The quantum theory of gravitons on an arbitrary background with a boundary has been discussed e.g. in Refs.~\cite{Espositobook, oneloop, gaugebound, moss13}. The boundary conditions must be gauge invariant in the same sense as discussed above for the Maxwell field. In a similar way as we showed above for $A_\mu$, it is shown that Dirichlet boundary conditions for all components of the field $h_{\mu\nu}$ and the ghost $\xi_\mu$ do not satisfy the gauge invariance requirement\footnote{A gauge transformation for the normal components $h_{\mu n}$ involves normal derivatives of $\xi_\mu$, which cannot be set to zero consistently with a Dirichlet boundary condition.}.  This already calls into question the validity of the conclusions of \cite{reuterbecker}, where in the context of the exact renormalization group for quantum gravity  different renormalizations are found for the bulk and boundary versions of Newton's constant, since there the calculation proceeds under the assumption of Dirichlet boundary conditions for all fields.

The analogy with the Maxwell field fails, however, in that been it has been proven that on a general background there are no gauge invariant mixed Robin-Dirichlet boundary conditions of the form we have been using; instead, all gauge invariant boundary 
conditions suitable for Lorentz (rotation) invariant gauge fixing (such as those introduced by Barvinsky \cite{barvinsky}) 
include tangential derivatives of  $h_{\mu\nu}$ at the boundary. This makes the heat-kernel ill-defined \cite{moss13, avramidi-ell}, rendering our whole calculational method inapplicable.

Nevertheless, there is a restricted class of backgrounds on which gauge invariant mixed Dirichlet-Robin boundary conditions can be found \cite{moss13, moss97}. They are characterized by the condition that the 
extrinsic curvature
of the boundary is proportional to the intrinsic boundary metric, with a proportionality coefficient which is constant over the boundary:
\begin{equation}
K_{ij} = \frac{K}{3}g_{ij}\,,\quad\quad \partial_i K=0\,.
\end{equation} 
where $i$ labels the tangential coordinates and $g_{ij}$ is the intrinsic boundary metric. On such backgrounds, and when the action is supplemented with the de Donder gauge-fixing term and the corresponding ghost term, the following mixed Dirichlet-Robin set of boundary conditions are gauge invariant\footnote{These boundary conditions are found in \cite{moss13}. They are quoted here with an error corrected in the first one ($2K^{ij}h_{ij}$ replacing $K g^{ij}h_{ij}$) and the sign of $K_{ij}$ is flipped everywhere to conform to our conventions.}:
\begin{subequations}\label{bcs}
\begin{align}
\nabla_n h_{nn}-K h_{nn}+2K^{ij}h_{ij}&=0\,,\label{hnn}\\
\nabla_n h_{ij} + K_{ij} h_{nn}&=0\,,\label{hij}\\
h_{in}= \xi_n&=0\,,\label{diri}\\
\nabla_n \xi_i+K_i^j\xi_j&=0\,.
\end{align}
\end{subequations}

Even though for the reasons discussed above we cannot assume the physical significance of the result, we will now outline the computation of the $a_2$ coefficient for these boundary conditions.
Since the wave operator for gravitons in the de Donder gauge is not of the Laplace type (\ref{laplace}), it is more convenient to switch variables from $h_{\mu\nu}$ to a scalar variable, the trace $\Phi$, and a tensor component, the traceless part $\hat{h}_{\mu\nu}=h_{\mu\nu}-\frac{1}{4}g_{\mu\nu}\Phi$. Then the quadratic Lagrangian decomposes into terms of the form $\Phi D^{(\Phi)}\Phi$, $\hat{h}_{\mu\nu}\,D^{(\hat{h})\,\,\mu\nu\,\rho\sigma}\,\hat{h}_{\rho\sigma}$ and $\bar{\xi}_\mu\, \tilde{D}^{\mu\nu}\,\xi_\nu$ for scalar, trace-free tensorial and ghost fluctuations, with each $D$ a Laplace-type operator\footnote{The exact form of the operators can be found in e.g. \cite{Duff}. The operators $D^{(\hat{h})\,\,\mu\nu\,\rho\sigma}$ and $\tilde{D}^{\mu\nu}$ are nonminimally coupled.}. In these variables the boundary conditions (\ref{bcs}) translate to:
\begin{subequations}\label{hatbcs}
\begin{align}
\nabla_n \Phi+ \frac{K}{2}\Phi  +2 K^{ij} \hat{h}_{ij} &= 0\,, \label{phibc} \\
\nabla_n \hat{h}_{ij}-\frac{1}{8}K_{ij} \Phi - \frac{3}{2} K_{ij} \,g^{kl}\hat{h}_{kl}&=0 \,,
\label{hatbc}\\
\hat{h}_{in}= \xi_n &=0\,,\\
\nabla_n \xi_i-K_i^j\xi_j&=0\,.\label{xii}
\end{align}
\end{subequations}
The relevant heat kernel coefficient should be computed as $a_2^{\mathrm{grav}}= a_2^{(\Phi)}+ a_2^{(\hat{h})} - 2 a_2^{(\xi)}$, using formula (\ref{a2}). The result is:
\begin{equation}\label{a2grav}
(96\pi^2)a_2^{\mathrm{grav}} = \int_{\mathcal{M}}\mathrm{d}^4x\,\sqrt{g}\, \left\{ -6R +20\Lambda_b \right\}+ 2\int_{\partial\mathcal{M}}\mathrm{d}^3x\,\sqrt{g}\, (- K)\,,
\end{equation}
Therefore the bulk-boundary balance fails to obtain. 

We now make several remarks about this calculation. 
\begin{enumerate}
\item[i)] 
Not only have we implicitly required the background to be on-shell (for the linearized theory to be gauge invariant), we have required it to have certain symmetry properties at the boundary. This makes even more unclear than already discussed above how the variation of the effective action with respect to $g_{\mu\nu}$ is supposed to proceed. 
\item[ii)] 
We have used Robin boundary conditions for the tangential components $h_{ij}$, but the balance of the Gibbons-Hawking term is required when varying the effective action and imposing Dirichlet boundary conditions for the intrinsic boundary metric. 
\item[iii)] The boundary conditions (\ref{bcs}) require the addition of a boundary action for $h_{\mu\nu}$, analogous to the one in (\ref{phiaction}), in order for the total action to be stationary at a solution to the equations of motion. If this boundary action does not vanish when the equations of motion are imposed (as happens for the scalar and the vector field), then the effective action can receive a further, boundary contribution that we have not considered. 
\item[iv)] 
The bulk-boundary balance in (\ref{a2grav}) is restored if 
$\Lambda_b$ is replaced by $R/4$, to which it is equal in a solution to the bare equation of motion.
This may be a mere coincidence, but perhaps it points to an important feature of a more consistent treatment of the problem.
\item[v)]  
The set of Robin boundary conditions for the scalar/traceless decomposition, and the heat kernel coefficient they imply, 
are new results, which could be of interest
quite apart from their application to studying the balanced variation of the effective action. 

\end{enumerate}

\section{Summary and discussion}

In this paper we have examined the one-loop renormalization of Newton's constant due to integrating out quadratic quantum fluctuations. This renormalization amounts to a quadratic divergence proportional to the $a_2$ heat kernel coefficient. This coefficient takes the general form 
\begin{equation}
a_2(D)\sim \beta_1 \int_{\mathcal{M}} \mathrm{d}^4x\,\sqrt{g}\,R\,+ \beta_2 \int_{\partial \mathcal{M}} \mathrm{d}^3x\,\sqrt{h}\,K\,
\end{equation}
 We have focused our attention on whether the two terms of this coefficient stand in the proper balance,
 \begin{equation}
 \beta_2 = 2\,\beta_1\,,
 \end{equation}
 that preserves the relationship of the bulk Einstein-Hilbert term to the boundary Gibbons-Hawking term in the effective action. This is necessary for the latter to yield the effective equations of motion upon variation with respect to the metric.

In Section 2 we established that this ``bulk-boundary balance'' depends on the interplay between the non-minimal coupling of the field and the boundary conditions imposed on the field fluctuations. For minimally coupled fields a Dirichlet or Neumann boundary condition produces the desired balance. On the other hand, 
we showed in section 3 that for non-minimally coupled fields 
the bulk-boundary balance is achieved only with the Robin boundary condition:
\begin{equation}\label{robinpref2}
\left(\nabla_n\varphi-\xi\,K\,\varphi\right)\big|_{\partial\mathcal{M}}=0\,.
\end{equation}
This boundary condition does not have any other justification we are aware of. (In particular, it is not conformally invariant for $\xi=1/6$.) It can be derived from an action including a particular boundary term, namely (\ref{phiaction}) with $\alpha=-1$, but this action is postulated \textit{ad hoc} for this purpose. One possibility is that if the nonminimal coupling emerges in an effective low-energy theory from integrating out minimally coupled degrees of freedom, as in \cite{onmodel}, this boundary action may emerge as well in the same way. This possibility requires further study.

In Section 4 we turned our attention to the Maxwell field, where the acceptable boundary conditions are restricted by the requirement of gauge invariance. The absolute and the relative boundary conditions, given respectively by (\ref{absolute}) and (\ref{relative}), are both gauge invariant. They are mixed Dirichlet-Robin boundary conditions, which include a Robin boundary condition similar to (\ref{robinpref2}) for either the normal or the tangential components of the vector potential,
and both preserve the bulk-boundary balance. This is the ``best-case scenario'', in which the boundary condition that induces the balanced renormalization has an independent justification, given here by the requirement of gauge invariance.

In Section 5 we examined the graviton field. We discussed a number of reasons why the 
question is not as straightforward to address as for the scalar and vector contributions,
and presented a set of gauge-invariant boundary conditions on a restricted class of backgrounds,
together with the resulting heat kernel coefficient. However, for the reasons explained in 
that section, we consider that analysis to be inconclusive. 

One possible resolution to the general problem of the bulk-boundary balanced renormalization is to allow the gravitational \textit{bare} action to be unbalanced, with a parameter $G_b$ in the bulk term and a different parameter $G_b^{(\partial)}$ in the boundary term. Then regardless of the boundary conditions it is possible to make the effective, renormalized values of these parameters coincide by fine-tuning the bare values. However, this procedure at most ``balances'' the effective action at a single renormalization scale, and hence is unsatisfactory in the context of the renormalization group, where we would desire the gravitational effective action to be well-defined and yield effective equations of motion at different values of the RG scale. Our results establish that this is possible only if the particular boundary conditions we have discussed are employed.

A further issue that should be addressed is whether
bulk/boundary balance should be, and is, maintained for the renormalization of the higher derivative terms in the effective 
action.\footnote{We thank S. Solodukhin for pointing this out.} 
This would be a more complicated problem to analyze, since the
higher derivatives would bring in more boundary terms and
more involved boundary conditions.

Finally, we remark on the connection between our results and the ``contact term'' that appears as part of the quantum correction to black hole entropy induced by nonminimally coupled fields \cite{kabat, solod3}. 
 In general, this quantum correction can be interpreted as a renormalization of Newton's constant as it appears in the Bekenstein-Hawking formula for the entropy \cite{suss,solod1}.
For example, for the nonminimally coupled scalar field the area term of the quantum-corrected entropy reads (cf. \ref{Grenscalar}):
\begin{equation}
S_{BH}=\frac{A}{4G_{\mathrm{0}}}= \frac{A}{4G_b}+\frac{A}{12\pi\epsilon^2}(1-6\xi)\,.
\end{equation}
Of the $1/\epsilon^2$ terms, the first one can be interpreted as the leading order divergence in the entanglement entropy across the horizon of the scalar fluctuations, while the second one (dubbed the ``contact term'') is a quantum contribution to the Noether charge, concentrated at the horizon, that appears for nonminimally coupled fields \cite{waldnonmin}. As discussed in \cite{reg1, satz}, the same result can be obtained in two ways: from an ``off-shell'' computation, where the partition function is computed on a space with a conical singularity, and from an ``on-shell'' computation, where 
the entropy comes from the Gibbons-Hawking boundary term of the effective action
evaluated at a smooth stationary point.

The equality between the two approaches, though, is achieved only if the bulk and boundary versions of Newton's constant renormalize in the same way. The off-shell computation matches inherently the renormalization of bulk $G$, while the on-shell computation matches that of boundary $G$. Naively, it would seem therefore that with Dirichlet boundary conditions (so that (\ref{Gbrenscalar}) with $\alpha=+\infty$ applies) the on-shell computation of black hole entropy yields a quantum correction consisting solely of the entanglement entropy, without the contact term. 

Nevertheless, this is not the correct conclusion to draw from our results. The starting point for the on-shell computation is a saddle-point evaluation of the gravitational thermal partition function on the solution of the effective equations of motion. If the bulk and boundary terms of the effective action are unbalanced, then the  
saddle point cannot be identified and the whole procedure is ill-defined. The actual conclusion is that the correct boundary conditions for all quantum fields (e.g.\ (\ref{robinpref2}), for a nonminimally coupled scalar) are necessary for any calculation that will involve treating gravity dynamically at some point. 
Once they are employed, the renormalization of boundary $G$ includes the nonminimal coupling, and the black hole entropy includes the contact term by whichever procedure it is computed.

\section{Acknowledgements}

We thank Cliff Burgess, William Donnelly, Martin Reuter,  Sergey Solodukhin and Matthew Williams for helpful discussions and comments.
This research was supported by NSF grant PHY-0903572.

\end{document}